\documentclass[final,3p,times]{elsarticle}

\usepackage{amssymb}

\journal{Physica A}

\begin{document}

\begin{frontmatter}



\title{The McLennan-Zubarev steady state distribution\\ 
and fluctuation theorems}


\author{Mitsusada M. Sano}

\address{ Graduate School of Human and Environmental Studies, \\
Kyoto University, Yoshida-Nihonmatsu-cho, Sakyo, Kyoto, 606-8501, Japan}

\begin{abstract}
The McLennan-Zubarev steady state distribution is studied 
in the connection with fluctuation theorems. 
We derive the McLennan-Zubarev steady state distribution 
from the nonequilibrium detailed balance relation. 
Then, considering the cumulant function or cumulant functional,  
two fluctuation theorems for entropy and for currents are proved. 
Using the fluctuation theorem for currents, 
the current is expanded in terms of thermodynamic forces. 
In the lowest order of the thermodynamic force, 
we find that the transport coefficient satisfies 
the Onsager's reciprocal relation. 
In the next order, we derived the correction term 
to the Green-Kubo formula.
\end{abstract}


\end{frontmatter}



\section{Introduction \label{sec1}}
Systems out equilibrium are filled in the nature, 
for instance, physical, chemical, and biological phenomena. 
In an isolated system, a nonequilibrium state spontaneously 
relaxes to thermodynamical equilibrium. 
On the other hand, a nonequilibrium steady state (NESS) is maintained 
or sustained by flows of energy and matter from the outside of a given system. 
Macroscopically, the fluxes and the entropy production are constant 
in the NESS. However, microscopically speaking, 
the fluxes and the entropy production fluctuate 
around macroscopic steady values. 
To elucidate the nature of these fluctuations is 
an issue of nonequilibrium statistical physics in last two decades, 
for instance, in refs.~\cite{EvansCohenMorriss93,GallavottiCohen95a,GallavottiCohen95b,Gallavotti96,Kurchan98,LebowitzSpohn99,Crooks99,Jarzynski00,EvansSearles02,Gaspard04,AndrieuxGaspard04,AndrieuxGaspard07a,AndrieuxGaspard07b}.

The entropy production is the most interesting quantity in the NESS, 
since it measures irreversibility of a given system. 
The fluctuation of the entropy production has a symmetry, 
that is, in the NESS, the logarithm of the probability ratio 
of the entropy production is related to the entropy production itself. 
\begin{equation}
\lim_{\tau\rightarrow \infty} \frac{1}{\tau} 
\ln \frac{P_{\tau}(\sigma)}{P_{\tau}(-\sigma)} = \sigma,
\label{eq:FT_0}
\end{equation}
where $\sigma$ is the entropy production and $P_{\tau}(\sigma)$ is 
the probability that the system exhibits the entropy production $\sigma$ 
in time interval $\tau$. 
This relation is called the fluctuation theorem, 
which was first found in numerical data of the simulations 
of the Nos\'{e}-Hoover thermostat systems\cite{EvansCohenMorriss93}. 
After that, a careful examination using dynamical systems theory 
and ergodic theory, has elucidated the phenomena 
observed\cite{GallavottiCohen95a,GallavottiCohen95b}. 
As a result, it is shown that the fluctuation of the entropy production is 
governed by time reversal symmetry of the system. 
The probability ratio in eq.~(\ref{eq:FT_0}) is 
a ratio of probabilities of forward path and its time reversed path. 
The fluctuation theorem is the consequence of the time reversal symmetry 
in the steady state. 
As another aspect, the fluctuation theorem affects transport phenomena. 
In ref.~\cite{Gallavotti96}, it was formulated 
as an extension of Onsager's reciprocity relation. 
Furthermore, the fluctuation theorem is not restricted to dynamical systems, 
and was also confirmed for the Langevin system\cite{Kurchan98}, 
for general stochastic systems\cite{LebowitzSpohn99}, 
and for the master equation\cite{Gaspard04,AndrieuxGaspard07b}. 
In ref.~\cite{LebowitzSpohn99}, for stochastic systems, 
the fluctuation theorem is rewritten 
in terms of the cumulant generating function $\mu(\lambda)$, 
i.e., $\mu(\lambda)= \mu(1-\lambda)$. 
We also call this symmetry the fluctuation theorem for the entropy production 
or Lebowitz-Spohn symmetry. 
Using the nonequilibrium detailed balance relation, 
microscopic derivations of the fluctuation theorem were attempted 
for a stochastic system\cite{Crooks99} 
and for a Hamiltonian system\cite{Jarzynski00}. 
For dynamical systems, the proof of the fluctuation theorem 
is still not in a satisfactory form except the case of 
the Gaussian thermostat systems or Nos\'{e}-Hoover systems. 
Here we emphasize that we treat the Hamiltonian dynamical systems, 
not stochastic systems.  We overcome deficiency of mathematical rigor 
in Jarzynski's treatment\cite{Jarzynski00}. 
As we shall see, the deficiency is due to the fact that he does not know 
the steady state distribution. 
In addition, our formulation 
is closely related to the work by Lebowitz and Spohn\cite{LebowitzSpohn99}.  
They do not derive the fluctuation theorem in the section 7, 
entitled "Stochastic and thermostatting heat reservoirs", 
of \cite{LebowitzSpohn99}. 
Our results may be what they tried to attempt.

Independent from the fluctuation theorem, 
the derivation of the steady state distribution 
in a form like a canonical ensemble was 
one of themes in nonequilibrium statistical physics. 
It was derived by McLennan\cite{McLennan59,McLennan63,McLennanBook90} and 
by Zubarev\cite{Zubarev62,ZubarevBook74}. 
Now their steady state distribution is called the McLennan-Zubarev 
steady state distribution. 
The McLennan-Zubarev steady state distribution was 
very formally derived. 
For stochastic systems (the master equation and the Langevin equation), 
the meaning of the McLennan-Zubarev steady state distribution 
is investigated 
by several authors\cite{Sano08,MaesNetocny10,ColangeliMaesWynats11}.
In these works, the steady state distribution near equilibrium is written 
in a similar form to the McLennan-Zubarev steady state distribution. 
Thus, the McLennan-Zubarev steady state distribution is considered 
as an approximation of the steady state distribution, tentatively. 
However, in recent years, 
in the considerations of 
the steady state thermodynamics\cite{KomatsuNakagawa08,KNST08,KNST09,KNST11}, 
the McLennan-Zubarev steady state distribution was derived 
through another route, i.e., the nonequilibrium detailed balance relation, 
which is closely related to the fluctuation theorem.  
In addition, the McLennan-Zubarev steady state distribution was 
also derived for quantum systems\cite{TasakiMatsui03,TasakiMatsui06}. 
These observations strongly stimulate us to reconsider 
the McLennan-Zubarev steady state distribution 
in a connection with the fluctuation theorem. 

The aim of this paper is to derive the McLennan-Zubarev steady 
state distribution from the nonequilibrium detailed balance relation, 
which is a key relation to derive the fluctuation theorem, 
and, in addition, to prove the fluctuation theorems 
for the entropy production and for the currents, 
by using the derived McLennan-Zubarev steady state distribution. 
Thus, we establish an exact connection 
between the McLennan-Zubarev steady 
state distribution and the fluctuation theorems. 
We confirm that the McLennan-Zubarev steady state distribution 
is exact, as far as we admit the nonequilibrium detailed balance relation.
For stochastic Markovian dynamics, the nonequilibrium detailed balance relation 
is justified\cite{Crooks98}. 
Unfortunately, however, 
we cannot justify the nonequilibrium detailed balance relation 
in our Hamiltonian formulation. This relation is an assumption more or less. 
We believe that the assumption of this relation holds. 
As shown later, 
the nonequilibrium detailed balance relation is strongly 
related not only to the McLennan-Zubarev steady state distribution,
but also to the fluctuation theorems.
The nonequilibrium detailed balance relation, eq.(53), implies 
all we need, i.e., McLennan-Zubarev steady state distribution and 
two fluctuation theorems.
As done for the master equation\cite{AndrieuxGaspard04,AndrieuxGaspard07a}, 
we obtain correction terms to the Green-Kubo formula, that is, 
the non-linear response, 
by using the cumulant generating functional for the currents. 
It is sure that the derived corrections are useful for investigations 
on the transport property.

This paper is organized as follows. 
In \S~\ref{sec2}, 
some known results on the McLennan-Zubarev steady state distribution 
are summarized. In \S~\ref{sec3}, 
after a preparation of the setting and notations, 
the McLennan-Zubarev steady state distribution is 
derived from the nonequilibrium detailed balance relation. 
In \S~\ref{sec4}, using a similar discussion to that in the previous section, 
the fluctuation theorem for the entropy production is proved.  
In \S~\ref{sec5}, the fluctuation theorem for the currents is proved. 
In \S~\ref{sec6}, the nonequilibrium detailed balance relation is argued. 
In \S~\ref{sec7}, we investigate the mean current. 
The mean current is expressed in a power series of thermodynamic forces. 
We derive the linear response (i.e., the Green-Kubo formula) 
and the non-linear response in the lowest order. 
As a by-product, we obtain non-trivial relations of 
the current correlation functions. 
These non-trivial relations are a kind of representations 
of time-reversal symmetry. 
In \S~\ref{sec8}, we summarize the results of this paper. 
\section{The McLennan-Zubarev steady state distribution \label{sec2}} 
In this section, we summarize a setting up for defining 
the McLennan-Zubarev steady state distribution.  
The basic facts on the McLennan-Zubarev steady state distribution 
are in \cite{ZubarevBook74} in details. 
The derivation of the McLennan-Zubarev steady state distribution, 
which is not the derivation in \cite{ZubarevBook74},  
will be shown in the next section. 

First, we summarize known results of nonequilibrium thermodynamics. 
We consider an isotropic one-component fluid. 
The entropy density $s$ satisfies the following entropy balance equation
\cite{deGrootMazur65,KitaharaYoshikawa94}.
\begin{equation}
\frac{\partial s}{\partial t} + \mbox{div}({\bf j}_{s}) = \sigma[s],
\end{equation}
where ${\bf j}_{s}$ is the entropy current
\begin{equation}
{\bf j}_{s}  = s{\bf u} + \frac{{\bf Q}}{T}  - \frac{\mu^{*}{\bf j}}{T}.
\label{eq:js}
\end{equation}
Here $T$ is the tempe\cite{LebowitzSpohn99}. rature. 
${\bf u}$ is the local fluid velocity. 
${\bf Q}$ is the heat current. 
${\bf j}$ is the diffusion current.
The first term of the right hand side of eq.(\ref{eq:js})
is the entropy current of the fluid.  
The second term is the entropy current of the energy flow.
The third term is the entropy current of the diffusion. 
$\mu$ is the chemical potential. We set 
\begin{equation}
\mu^{*} = \mu - \frac{{\bf u}^{2}}{2}.
\end{equation}
The energy current ${\bf j}_{\varepsilon}$  is given by 
\begin{equation}
{\bf j}_{\varepsilon}  = {\bf Q} - {\bf u}:\sigma'.
\end{equation}
$\sigma'$ is the viscous stress tensor. 
The stress tensor $\mathsf{T}$ is related to 
the viscous stress tensor as follows.
\begin{equation}
\mathsf{T}_{ij} = - P \delta_{ij} + \sigma_{ij}'.
\end{equation}
$P$ is the hydrostatic pressure. 
The product of vector and tensor ${\bf u}:\sigma'$ is 
\begin{equation}
({\bf u}:\sigma')_{i} = \sum_{j}u_{j}\sigma'_{ji}.
\end{equation}
The entropy production rate is given by 
\begin{equation}
\sigma[s] = {\bf j}_{\varepsilon} \cdot \nabla \left ( \frac{1}{T}\right )
+ (-\sigma') : \nabla \left ( -\frac{{\bf u}}{T}\right )
+ {\bf j} \cdot \nabla \left ( -\frac{\mu^{*}}{T}\right ).
\label{eq:epr}
\end{equation}
The notation of the tensor product is 
\begin{equation}
(-\sigma') : \nabla \left ( -\frac{{\bf u}}{T}\right ) = 
\sum_{i,j} (-\sigma'_{ij}) 
\frac{\partial}{\partial x_{j}}\left ( -\frac{u_{i}}{T} \right ).
\end{equation}
From the expression of $\sigma[s]$, 
we see that the entropy production rate is 
(the irreversible part of the current of the conserved quantity)
$\times$ (the gradient of the intensive parameter).
Each term is related to the energy conservation law,  
the momentum conservation law, and the particle number conservation law. 
This fact is important when we consider the entropy production rate 
of a given system not only in a macroscopic description, 
but also microscopic description. 

Now we turn to microscopic description of a one-component fluid system. 
Consider a system with $N$ point particles with mass $m$. 
The interaction potential is given by $V(|{\bf q}_{i}-{\bf q}_{j}|)$. 
The Hamiltonian is 
\begin{equation}
H = \sum_{n=1}^{N} \frac{{\bf p}_{i}^{2}}{2m} + 
\sum_{i<j}^{N} V(|{\bf q}_{i}-{\bf q}_{j}|).
\label{eq:Hamiltonian}
\end{equation}
The equations of motion is 
\begin{equation}
\frac{d{\bf q}_{i}}{dt} = \frac{\partial H}{\partial {\bf p}_{i}},\;\;
\frac{d{\bf p}_{i}}{dt} = -\frac{\partial H}{\partial {\bf q}_{i}}.
\label{eq:eq-motion}
\end{equation}
In this paper, we use the following notation for the variables of 
position and momentum. 
$\Gamma=({\bf q}_{1},{\bf q}_{2},\dots,{\bf q}_{N},
{\bf p}_{1},{\bf p}_{2},\dots,{\bf p}_{N})$. 
The Hamiltonian $H$ is a function of $\{{\bf q}_{i}\}_{i=1}^{N}$ and 
$\{{\bf p}_{i}\}_{i=1}^{N}$. 
The Hamiltonian may written as 
$H=H(\{{\bf q}_{i}\},\{{\bf p}_{i}\})$.
Symmetry which the system possesses is important. 
This Hamiltonian has a symmetry. 
\begin{equation}
H(\{{\bf q}_{i}\},\{{\bf p}_{i}\})=H(\{{\bf q}_{i}\},-\{{\bf p}_{i}\})
\label{eq:HamiltonianSymmetry}. 
\end{equation}
Time-reversal symmetry of orbits can be written as
\begin{equation}
{\bf q}_{i,\mbox{\scriptsize R}}(t) = {\bf q}_{i}(-t),\;\; 
{\bf p}_{i,\mbox{\scriptsize R}}(t) = -{\bf p}_{i}(-t).
\end{equation}
For the orbit $\Gamma(t)$, 
the time-reversed orbit is written as $\Gamma_{\mbox{\scriptsize R}}(t)$. 

This system has three conservation laws. 
(1)Particle number conservation law:
\begin{equation}
\frac{\partial n(\Gamma,{\bf x})}{\partial t} + 
\mbox{div} ({\bf j}(\Gamma,{\bf x})) = 0.
\end{equation}
$n(\Gamma,{\bf x})$ is the particle density. 
${\bf j}(\Gamma,{\bf x})$ is the current density of particles.
(2)Energy conservation law: 
\begin{equation}
\frac{\partial H(\Gamma,{\bf x})}{\partial t} + 
\mbox{div}({\bf J}_{H}(\Gamma,{\bf x})) = 0.
\end{equation}
$H(\Gamma,{\bf x})$ is the energy density. 
$J_{H}(\Gamma,{\bf x})$ is the energy current density. 
(3)Momentum conservation law: 
\begin{equation}
\frac{\partial {\bf p}(\Gamma,{\bf x})}{\partial t} + 
\mbox{div}(\mathsf{T}(\Gamma,{\bf x})) = 0.
\end{equation}
$\mathsf{T}(\Gamma,{\bf x})$ is the momentum tensor. 
Here the functions appeared in the above conservation laws 
are the density of particles, 
\begin{equation}
n(\Gamma,{\bf x}) = \sum_{i=1}^{N} \delta({\bf q}_{i}-{\bf x}),
\end{equation}
the momentum density,
\begin{equation}
{\bf p}(\Gamma,{\bf x}) = 
\sum_{i=1}^{N} {\bf p}_{i} \delta({\bf q}_{i}-{\bf x}),
\end{equation}
and the current of particles,
\begin{equation}
{\bf j}(\Gamma,{\bf x};t) = 
\sum_{i=1}^{N} \frac{{\bf p}_{i}}{m} \delta({\bf q}_{i}-{\bf x}).
\end{equation}
In order to capture heat, 
we define the energy density is given by 
\begin{eqnarray}
H(\Gamma,{\bf x}) & = & 
\sum_{i=1}^{N} \left \{ \frac{{\bf p}_{i}^{2}}{2m} + 
\frac{1}{2} \sum_{j\neq i}V(|{\bf q}_{i}-{\bf q}_{j}|) \right \} 
\delta({\bf q}_{i}- {\bf x}).
\end{eqnarray}
For heat current, 
the energy current is defined as 
\begin{eqnarray}
{\bf J}_{H}(\Gamma,{\bf x};t) & = & 
\sum_{i}^{N}\left [ \frac{{\bf p}_{i}^{2}}{2m} + 
\frac{1}{2} \sum_{j\neq i} V(|{\bf q}_{i}-{\bf q}_{j}|)\right ] 
\frac{{\bf p}_{i}}{m}\delta({\bf q}_{i}-{\bf x}) 
+ \frac{1}{4m} \sum_{j\neq i} (({\bf p}_{i}+{\bf p}_{j})\cdot{\bf F}_{ij})
({\bf q}_{i}-{\bf q}_{j}) \delta({\bf q}_{i}-{\bf x}).
\end{eqnarray}
The momentum tensor $\mathsf{T}(\Gamma,{\bf x};t)$ is 
\begin{eqnarray}
\mathsf{T}_{\beta\alpha}(\Gamma,{\bf x};t) & = & 
\frac{1}{2}\sum_{i\neq j} ({\bf q}_{i}-{\bf q}_{j})_{\beta}(F_{ij})_{\alpha}
\delta({\bf q}_{i}-{\bf x}) 
+ \frac{1}{m}\sum_{i=1}^{N}({\bf p}_{i})_{\beta}({\bf p}_{i})_{\alpha}
\delta({\bf q}_{i}-{\bf x}),
\end{eqnarray}
Here the force is 
\begin{equation} 
{\bf F}_{ij}= - \frac{\partial}{\partial {\bf q}_{i}}
V(|{\bf q}_{i}-{\bf q}_{j}|).
\end{equation}
${\bf F}_{ij}({\bf q}_{i}-{\bf q}_{j})$ is the force, 
which the $j$-th particle acts to the $i$-th particle

As a result, the above three conservation law implies 
three phenomena, i.e.,  thermal conduction, momentum diffusion, and diffusion.
Here we define three currents corresponding to the 
thermal conduction, momentum diffusion, and diffusion.
\begin{eqnarray}
{\bf j}_{0}(\Gamma,{\bf x};t) & = & {\bf J}_{H}(\Gamma,{\bf x};t) \\
{\bf j}_{1}(\Gamma,{\bf x};t) & = & \mathsf{T}(\Gamma,{\bf x};t) \\
{\bf j}_{2}(\Gamma,{\bf x};t) & = & {\bf j}(\Gamma,{\bf x};t). 
\end{eqnarray}
Next we define three quantities. 
\begin{eqnarray}
P_{0}(\Gamma,{\bf x})& = & H(\Gamma,{\bf x}) \\
P_{1}(\Gamma,{\bf x})& = & {\bf p}(\Gamma,{\bf x}) \\
P_{2}(\Gamma,{\bf x})& = & n(\Gamma,{\bf x}).
\end{eqnarray}
Here $\mu^{*}$ is the chemical potential, which is defined as  
\begin{equation}
\mu^{*} = \mu - \frac{m{\bf u}^{2}}{2},
\end{equation}
where ${\bf u}$ is the velocity field.
If the velocity field exists, the chemical potential would be 
subtracted by the amount of the velocity field. 
$\beta({\bf x})$ is the inverse temperature divided by the Boltzmann constant. 
\begin{equation}
\beta({\bf x}) = \frac{1}{k_{\mbox{\scriptsize B}}T({\bf x})},
\end{equation}
Here $T({\bf x})$ is the temperature field. 
We also define the following quantities.
\begin{eqnarray}
F_{0}({\bf x}) & = & \beta({\bf x}) \\
F_{1}({\bf x}) & = & -\beta({\bf x}) {\bf u}({\bf x})\\
F_{2}({\bf x}) & = & -\beta({\bf x}) \mu^{*}({\bf x}).
\end{eqnarray}
The thermodynamic forces ${\bf X}_{i}=\nabla F_{i}$ are defined as
\begin{eqnarray}
{\bf X}_{0}({\bf x})& = & \nabla\beta({\bf x}) \\
{\bf X}_{1}({\bf x})& = & -\nabla(\beta({\bf x}){\bf u}) \\
{\bf X}_{2}({\bf x})& = & -\nabla(\beta({\bf x}) \mu^{*}).
\end{eqnarray}
For convenience, we define a vector, which combines three  
thermodynamic forces. 
\begin{equation}
{\bf X}({\bf x}) = 
\left (
\begin{array}{c}
{\bf X}_{0}({\bf x})\\
{\bf X}_{1}({\bf x})\\
{\bf X}_{2}({\bf x})
\end{array}
\right ).
\end{equation}

In nonequilibrium thermodynamics, 
the local entropy production rate $\sigma[s]$ is 
defined as eq.(\ref{eq:epr}). 
Now we define the microscopic local entropy production rate 
$\sigma[\Gamma,{\bf x}]$ for our particle system. 
\begin{eqnarray}
\sigma[\Gamma,{\bf x}]
& = & \sum_{m} {\bf j}_{m}(\Gamma,{\bf x};t) 
\cdot {\bf X}_{m}({\bf x}) \nonumber \\
& = & {\bf J}_{H}(\Gamma,{\bf x};t)\cdot \nabla \beta({\bf x}) 
+ \mathsf{T}(\Gamma,{\bf x};t):\nabla(- \beta({\bf x}) {\bf u})
+ {\bf j}(\Gamma,{\bf x};t)\cdot\nabla( - \beta({\bf x}) \mu^{*} ).
\label{eq:epr0}
\end{eqnarray}
Precisely speaking, this quantity is the local entropy production rate 
divided by the Boltzmann constant $k_{B}$. 
Equation (\ref{eq:epr0}) is a first key assumption in this paper. 
Other assumptions are the nonequilibrium detailed balance relation, 
and the time-reversal symmetry, which will be explained later. 
These three assumptions are the key relations in this paper. 

The local equilibrium state is characterized 
by a local equilibrium distribution. 
That is 
\begin{equation}
\rho_{\ell}(\Gamma) = Z_{0}^{-1} 
\exp \left [ - \sum_{m} \int d{\bf x} \;
F_{m}({\bf x}) P_{m}(\Gamma,{\bf x})
\right ],
\label{eq:local_eq}
\end{equation}
where 
\begin{equation}
Z_{0} = \int d\Gamma \exp \left [ - \sum_{m} \int d{\bf x} \;
F_{m}({\bf x}) P_{m}(\Gamma,{\bf x})
\right ].
\end{equation}
$\ell$ stands for local equilibrium. 
This distribution does not describe a steady state. 
To obtain the steady state distribution, 
we should treat the deviation from the local equilibrium. 
The deviation includes the time-integral of the entropy production rate.

Here we ``simply'' write down the McLennan-Zubarev steady state distribution 
which appeared in the book of Zubarev\cite{ZubarevBook74}. 
By the above setting, the McLennan-Zubarev steady state distribution is given 
as follows\cite{McLennan59,McLennan63,McLennanBook90,Zubarev62,ZubarevBook74}.
\begin{eqnarray}
\rho_{\mbox{\scriptsize ss}}(\Gamma) 
& = & Z^{-1} 
\exp \left [ -\sum_{m} \int d{\bf x}\; F_{m}({\bf x})P_{m}(\Gamma,{\bf x})
+ \sum_{m}\int d{\bf x} \int_{-\infty}^{0} dt\; 
{\bf j}_{m}(\Gamma,{\bf x};t)\cdot {\bf X}_{m}({\bf x}) \right ]. 
\label{eq:MZdist}
\end{eqnarray}
``ss'' means the steady state. 
Here $\Gamma$ is 
$\Gamma = ({\bf q}_{1},{\bf q}_{2},\dots,{\bf q}_{N},
{\bf p}_{1},{\bf p}_{2},\dots,{\bf p}_{N})$. 
$Z$ is the normalization constant. 
\begin{eqnarray}
Z 
& = & 
\int d\Gamma\;  
\exp \left [ -\sum_{m} \int d{\bf x}\; F_{m}({\bf x})P_{m}(\Gamma,{\bf x})
+ \sum_{m}\int d{\bf x} \int_{-\infty}^{0} dt\; 
{\bf j}_{m}(\Gamma,{\bf x};t)\cdot {\bf X}_{m}({\bf x}) \right ].
\end{eqnarray}
${\bf j}_{m}(\Gamma,{\bf x};t)$ is the energy current density $(m=0)$, 
the momentum current density $(m=1)$,
and the current density of the particle number $(m=2)$. 
It might be that an explanation on the notations is needed here. 
The coordinates of $\Gamma$ and ${\bf x}$ 
is the coordinate of the orbits and of the density, respectively.  
${\bf j}_{m}(\Gamma,{\bf x};t)$ is the density of the current. 
The McLennan-Zubarev steady state distribution is near
the local equilibrium state, eq.(\ref{eq:local_eq}). 
The deviation from the local equilibrium is
\begin{equation}
\sum_{m}\int d{\bf x} \int_{-\infty}^{0} dt \; 
{\bf j}_{m}(\Gamma,{\bf x};t)\cdot {\bf X}_{m}({\bf x}).
\end{equation}
This term of eq.~(\ref{eq:MZdist}) expresses 
the total amount of the entropy production for a given orbit. 
$\rho_{\mbox{\scriptsize ss}}(\Gamma)$ has a physical meaning 
that at infinite past ($t=-\infty$), the system is in
a local equilibrium. Then after, up to the present ($t=0$), 
the system produces the whole of the entropy production. 
This amount of the total entropy production is in the exponential function. 

\section{Derivation of the McLennan-Zubarev steady state  
distribution \label{sec3}} 
In this section, we derive the McLennan-Zubarev steady state distribution 
from the nonequilibrium detailed balance relation. 
Assumptions are needed to derive the McLennan-Zubarev 
steady state distribution:  
(a) the entropy production rate, eq.(\ref{eq:epr0}), 
(b) the time-reversibility, eq.(\ref{eq:TRsymmetry}), and 
(c) the nonequilibrium detailed balance relation, eq.(\ref{eq:NDBR}). 
The nonequilibrium detailed balance relation is needed 
to derive the McLennan-Zubarev steady state distribution and 
the fluctuation theorems. 
The time reversibility is essential to the fluctuation theorems. 
Before starting the derivation, 
we prepare a setting and notations for later use. 
\subsection{Preparation\label{sec3-1}}
In this subsection, we present the notations and
the time-reversal symmetry,
which are important for our formulation.
Here we do not follow McLennan's derivation, 
but will present another derivation 
by using the nonequilibrium detailed balance relation.
The McLennan's derivation is presented in Appendix A for comparison.

\begin{figure}
\begin{center}
\includegraphics[width=12cm]{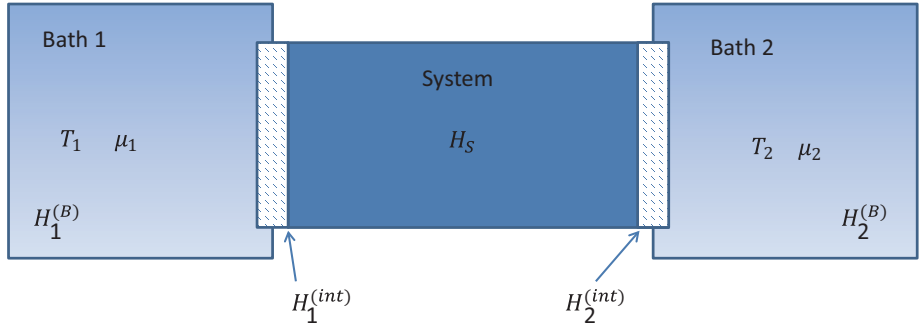}
\caption{\label{fig1}
The configuration of the system and the baths: 
The system is sandwiched between the baths 1 and 2, 
whose temperatures and chemical potentials are different 
($T_{1} \neq T_{2}$, $\mu_{1} \neq \mu_{2}$). 
Hamiltonians $H_{S}$, $H_{i}^{(B)}$, and $H_{i}^{(int)}$ 
($i=1,2$) describe 
the system, the $i$-th bath, and the interaction between the system 
and the $i$-th bath, respectively
}
\end{center}
\end{figure}
Consider a system which consists of particles. 
The total system has the system and two baths.
The Hamiltonian of the total system is
\begin{equation}
H_{tot} = H_{S} + H_{B} + U.
\label{eq:hamiltonian}
\end{equation}
$H_{S}$ is the Hamiltonian of the system. 
$H_{B}$ is the Hamiltonian of the baths.
$U$ is the interaction between the system and the baths.
The configuration of the whole system is depicted 
in Fig.\ref{fig1}. 
The bath $i$ is in equilibrium and has temperature $T_{i}$ ($i=1,2$), 
and has chemical potential $\mu_{i}$ ($i=1,2$).
The system (the bath $i$) has $N_{s}$ ($N_{i}$) particles, respectively,
$N = N_{s} + N_{1} + N_{2}$. 
The total system has $N$ particles. All $N$ particles are identical. 
Imagine that the system is under the heat conduction 
and the particle flow with this nonequilibrium boundary condition. 
We define the Hamiltonian of the total system as follows.
\begin{equation}
H_{S} = H_{S}(\Gamma_{s}),
\end{equation}
\begin{equation}
H_{B} = \sum_{i=1}^{2} H_{i}^{(B)}(\Gamma_{b}^{(i)}),
\end{equation}
\begin{equation}
U = \sum_{i=1}^{2} H_{i}^{(int)}(\Gamma_{b}^{(i)},\Gamma_{s}).
\end{equation}
$\Gamma_{s}$ is the coordinates and momenta of the system. 
$\Gamma_{b}^{(i)}$ is the coordinates and momenta of the bath $i$. 
$H_{i}^{(B)}$ is the Hamiltonian of the bath $i$. 
$H_{i}^{(int)}$ represents the interaction between the system and the bath $i$. 
The state of the system and the baths is determined 
by the coordinates and the momenta. 
\begin{equation}
\Gamma_{s} = 
({\bf q}_{1,s},\cdots,{\bf q}_{N_{s},s}
,{\bf p}_{1,s},\cdots,{\bf p}_{N_{s},s}), 
\end{equation}
\begin{equation}
\Gamma_{b}^{(i)} = 
({\bf q}_{1,i},\cdots,{\bf q}_{N_{i},i}
,{\bf p}_{1,i},\cdots,{\bf p}_{N_{i},i}). 
\end{equation}
The Hamiltonians are defined as
\begin{equation}
H_{S}(\Gamma_{s}) = 
\sum_{j=1}^{N_{s}} \frac{|{\bf p}_{j,s}|^{2}}{2m}
+ V_{s}({\bf q}_{1,s},\cdots,{\bf q}_{N_{s},s}),
\end{equation}
\begin{equation}
H_{i}^{(B)}(\Gamma_{b}^{(i)}) = 
\sum_{j=1}^{N_{i}} \frac{|{\bf p}_{j,b}|^{2}}{2m}
+ V_{i}({\bf q}_{1,i},\cdots,{\bf q}_{N_{i},i}),
\end{equation}
\begin{equation}
H_{i}^{(int)}(\Gamma_{b}^{(i)},\Gamma_{s}) = 
U_{i,s}({\bf q}_{1,i},\cdots,{\bf q}_{N_{i},i},
{\bf q}_{1,s},\cdots,{\bf q}_{N_{s},s}).
\end{equation}
We write $d\Gamma_{s}=\prod_{i=1}^{N_{s}}d{\bf q}_{i,s}d{\bf p}_{i,s}$ 
and $d\Gamma_{b}^{(i)}=\prod_{j=1}^{N_{i}}d{\bf q}_{j,i}d{\bf p}_{j,i}$. 
It is clear that the system is driven out of equilibrium 
with the nonequilibrium boundary conditions $T_{1}\neq T_{2}$ 
and $\mu_{1}\neq \mu_{2}$. 
In the system, the heat conduction and the particle flow are achieved.

The time-reversal operation is essential 
to the consideration of the fluctuation theorem. 
Flipping the sign of the momenta is a basic operation. 
We define the $^{*}$-operation by 
\begin{equation}
\Gamma_{s}^{*} = ({\bf q}_{1,s},{\bf q}_{2,s},\dots,{\bf q}_{N_{s},s},
-{\bf p}_{1,s},-{\bf p}_{2,s},\dots,-{\bf p}_{N_{s},s}).
\end{equation}
The ``whole'' system considered in this paper 
is a Hamiltonian dynamical system. 
But the system is a certain dynamical system with dissipation 
by the existence of the baths or reservoirs. 
The Hamiltonian does not contain the magnetic field and 
has the symmetry of eq.~(\ref{eq:HamiltonianSymmetry}). 
We denote the whole phase space by ${\cal S}$. 
${\cal S}$ is 
${\cal S}= {\cal S}_{\mbox{\scriptsize c}}^{s} \times {\cal S}_{\mbox{\scriptsize m}}^{s}
\times 
{\cal S}_{\mbox{\scriptsize c}}^{b(1)} \times {\cal S}_{\mbox{\scriptsize m}}^{b(1)} 
\times 
{\cal S}_{\mbox{\scriptsize c}}^{b(2)} \times {\cal S}_{\mbox{\scriptsize m}}^{b(2)} 
= {\cal S}_{s} \times {\cal S}_{b}$,
where ${\cal S}_{\mbox{\scriptsize c}}^{s}$ (${\cal S}_{\mbox{\scriptsize c}}^{b(i)}$)
is the coordinate space of the system (the bath $i$), respectively.
and ${\cal S}_{\mbox{\scriptsize m}}^{s}$ (${\cal S}_{\mbox{\scriptsize m}}^{b(i)}$)
is the momentum space of the system (the bath $i$), respectively. 
We denote the flow of this whole system by $\Phi$. 
If we use the notation of $\Gamma(t)=(\Gamma_{s},\Gamma_{b}^{(1)},
\Gamma_{b}^{(2)})$ 
for the state, the flow acts as
\begin{equation}
\Gamma(t+\tau) = \Phi^{\tau} \Gamma(t). 
\end{equation}
We can write the time-evolution of the distribution 
using the Perron-Frobenius operator.
\begin{equation}
\rho(\gamma; t+\tau) = \int d\Gamma \;
\delta( \Phi^{\tau}\Gamma - \gamma ) 
\rho(\Gamma;t).
\end{equation}
We call the delta function in the above equation 
the integral kernel of the Perron-Frobenius operator. 
This type of delta functions will be frequently used later.  
In this paper, we may fix $\tau$ in $\Phi^{\tau}$. 
Then, after some manipulations, 
we shall take the limit $\tau\rightarrow \infty$. 
Thus, it is convenient to define the map ${\cal T}$ as ${\cal T}=\Phi^{\tau}$. 
By the map which the whole map ${\cal T}$ is restricted to the system, 
the phase space volume element of the system contracts. 

The map ${\cal T}$ satisfies the following time reversal symmetry.
\begin{equation}
({\cal T}(\Gamma))^{*} = {\cal T}^{-1}(\Gamma^{*}),
\label{eq:TRsymmetry}
\end{equation}
for any $\Gamma \in {\cal S}$. 
This symmetry will be used later 
and is the second key relation 
to derive the McLennan-Zubarev steady state distribution 
and to prove the fluctuation theorems.

\subsection{Derivation\label{sec3-2}}
We denote the entropy production contribution 
from a specific path $\nu \rightarrow \gamma$, 
by $\Sigma[\nu \rightarrow \gamma]$, 
where $\nu, \gamma \in {\cal S}_{s}$. 
This quantity is given by 
\begin{equation}
\Sigma[\nu\rightarrow \gamma] = 
\sum_{m} 
\int d{\bf x} \int_{-\tau}^{0} dt\; 
{\bf j}_{m}(\Gamma_{s},{\bf x};t)\cdot {\bf X}_{m}({\bf x}), 
\end{equation}
where $\gamma = ({\cal T}(\Gamma))_{s}$ with $\Gamma_{s} = \nu$.
The integral of ${\bf x}$ is taken over the system. 
Hereafter without specifying, $\int d{\bf x}$ represents 
the integral over the system, not over the whole system. 
The third key relation in our derivation is 
the nonequilibrium detailed balance relation.  
For various derivations of the fluctuation theorem, 
the nonequilibrium detailed balance relation was used\cite{Crooks99}. 
The nonequilibrium detailed balance relation for our case is given by  
\begin{equation}
P[\nu\rightarrow\gamma] = 
e^{\Sigma[\nu\rightarrow \gamma]} P[\gamma^{*}\rightarrow\nu^{*}],
\label{eq:NDBR}
\end{equation}
where $P[\nu\rightarrow\gamma]$ is the kernel of the 
Perron-Frobenius operator which casts 
from the state $\nu$ to the state $\gamma$. 
Since we consider a dynamical system, 
the function $P[\nu\rightarrow\gamma]$ 
is given by a delta function. 
\begin{equation}
P[\nu\rightarrow\gamma] = \delta (({\cal T}(\Gamma))_{s}-\gamma).
\end{equation}
Equation (\ref{eq:NDBR}) is our assumption, 
since we cannot derive eq.~(\ref{eq:NDBR}) 
from the equations of motion by first principle. 
However, the nonequilibrium detailed balance relation gives us 
the connection between entropy production rate and 
the stability of the system 
\cite{GallavottiCohen95a,GallavottiCohen95b}. 
This can be easily confirmed by rewriting $\delta$-function 
and integrating eq.~(\ref{eq:NDBR}) with respect to $\nu=\Gamma_{s}$. 
We get 
\begin{equation}
\left | \frac{\partial ({\cal T}(\Gamma))_{s}}{\partial \Gamma_{s}} \right |^{-1}
= e^{\Sigma[\nu\rightarrow \gamma]}.
\end{equation}
Thus, the entropy production is directly related to 
the stability of a given system.  
Here we set $\nu = \Gamma_{s}$, $\Gamma'={\cal T}(\Gamma)$, 
$\gamma = (\Gamma')_{s}$, 
and $\Gamma''= {\cal T}^{-1}(\Gamma)$.

Let us consider time evolution of a given state. 
We assume that the initial state is $\nu$ and 
the initial distribution is in a canonical state, 
i.e., a local equilibrium state. 
This can be realized by defining the distribution $P_{\nu}(\Gamma_{s})$ 
\begin{equation}
P_{\nu}(\Gamma_{s}) = \rho_{\ell}(\Gamma_{s}),
\end{equation}
where
\begin{equation}
\rho_{\ell}(\Gamma) = 
\exp \left [ -\int d{\bf x} \left \{ 
\beta({\bf x}) H(\Gamma, {\bf x}) 
- \beta({\bf x}){\bf u}({\bf x}) \cdot {\bf p}(\Gamma,{\bf x}) 
- \beta({\bf x}) n(\Gamma,{\bf x}) \mu^{*}({\bf x})
\right \}
\right].
\end{equation}
In $P_{\nu}(\Gamma_{s})$, $\nu$ represents the initial state. 
After time evolution during time interval $\tau$, the system will be found 
in the state $\gamma \in {\cal S}_{s}$ with the probability 
\begin{equation}
\rho_{\nu}(\gamma) = \int d\Gamma_{s}\; P_{\nu}(\Gamma_{s}) 
\delta(({\cal T}(\Gamma))_{s}-\gamma).
\end{equation}

Let us start the derivation of the McLennan-Zubarev steady state distribution.
Using the nonequilibrium detailed balance relation, eq.~(\ref{eq:NDBR}), 
we get 
\begin{eqnarray}
\rho_{\nu}(\gamma) 
& = & 
\int d\Gamma_{s} \; P_{\nu}(\Gamma_{s}) P[\nu\rightarrow \gamma] \nonumber \\
& = & 
\int d\Gamma_{s} \; P_{\nu}(\Gamma_{s}) e^{\Sigma[\nu\rightarrow \gamma]}
P[\gamma^{*}\rightarrow \nu^{*}] \nonumber \\
& = & 
\int d\Gamma_{s} \; P_{\nu}(\Gamma_{s}) e^{\Sigma[\nu\rightarrow \gamma]}
\delta(({\cal T}({\cal T}(\Gamma)^{*}))_{s} - \nu^{*}).
\label{eq:MZ_NEDBR}
\end{eqnarray}
Here we note that ${\cal T}(\Gamma)^{*} = ({\cal T}^{-1}(\Gamma^{*}))$.
Thus, we obtain
\begin{eqnarray}
\rho_{\nu}(\gamma) 
& = & 
\int d\Gamma_{s} \; \rho_{\ell}(\Gamma_{s}) 
e^{\Sigma[\nu\rightarrow \gamma]} \delta(\Gamma_{s}^{*}-\nu^{*}) 
\nonumber \\
& = & 
\int d\Gamma_{s} \; \rho_{\ell}(\Gamma_{s}) 
e^{\Sigma[\nu\rightarrow \gamma]} \delta(\Gamma_{s}-\nu) 
\nonumber \\
& = & 
\rho_{\ell}(\nu) e^{\Sigma[\nu\rightarrow \gamma]} .
\end{eqnarray}
We note that taking the limit $\tau \rightarrow \infty$, 
the steady state is obtained as   
\begin{equation}
\rho_{\mbox{\scriptsize ss}}(\gamma)
= \lim_{\tau\rightarrow \infty} Z_{\tau}^{-1} \rho_{\nu}(\gamma).
\label{eq:ss}
\end{equation}
where $Z_{\tau}$ is the normalization constant, which is given by
\begin{eqnarray}
Z_{\tau}
& = &
\int d\Gamma_{s}\; 
\exp \left [ -\sum_{m} \int d{\bf x}\; F_{m}({\bf x})P_{m}(\Gamma_{s},{\bf x})
+ \sum_{m}\int d{\bf x} \int_{-\tau}^{0} dt\; 
{\bf j}_{m}(\Gamma_{s},{\bf x};t)\cdot {\bf X}_{m}({\bf x}) \right ].
\end{eqnarray}
Inserting the expression for $\rho_{\nu}(\gamma)$ into Eq.(\ref{eq:ss}), 
we find
\begin{eqnarray}
\rho_{\mbox{\scriptsize ss}} (\gamma)
& = & 
\lim_{\tau\rightarrow\infty} Z_{\tau}^{-1} \rho_{\ell}(\nu) 
e^{\Sigma[\nu\rightarrow \gamma]} .
\label{eq:MZdist_derived}
\end{eqnarray}
Thus, from the nonequilibrium detailed balance relation, eq.~(\ref{eq:NDBR}), 
we have derived the McLennan-Zubarev steady state distribution, 
eq.~(\ref{eq:MZdist}). 
\section{Fluctuation theorem for the entropy production \label{sec4}} 
In this section, we prove the fluctuation theorem for the entropy production 
as a symmetry relation of a cumulant generating function.  
Now define the cumulant generating function $\mu(\lambda)$. 
\begin{equation}
\mu(\lambda) = 
\lim_{\tau\rightarrow \infty} -\frac{1}{\tau} 
\ln \left \langle 
\exp \left [
- \lambda \sum_{m} \int d{\bf x} \int_{0}^{\tau} dt\; 
{\bf j}_{m}(\Gamma_{s},{\bf x};t)\cdot {\bf X}_{m}({\bf x})
\right ]
\right \rangle_{\mbox{\scriptsize ss}}.
\label{eq:mu_lambda}
\end{equation}
Here $\langle \cdots \rangle_{\mbox{\scriptsize ss}}$ is the average 
over the McLennan-Zubarev steady state distribution. 
Note that the McLennan-Zubarev steady state distribution 
includes the normalization constant $Z$. 
We shall prove that this cumulant generating function has 
the following symmetry, 
which Lebowitz and Spohn discovered 
for stochastic systems\cite{LebowitzSpohn99}, 
\begin{equation}
\mu(\lambda) = \mu(1-\lambda).
\end{equation}
We call this relation the Lebowitz-Spohn symmetry. 

First, we define the probability 
that the entropy production is $X$ along a path $\nu\rightarrow \gamma$ 
during the time interval $\tau$ by $p_{\tau}^{\mbox{\scriptsize F}}(X;\gamma)$.
\begin{equation}
p_{\tau}^{\mbox{\scriptsize F}}(X;\gamma)
 = \int d\Gamma_{s}\; P_{\nu}(\Gamma_{s}) \delta(({\cal T}(\Gamma))_{s}-\gamma) 
\delta(X-\Sigma[\Gamma_{s}\rightarrow({\cal T}(\Gamma))_{s}]).
\label{eq:PtauXnugamma}
\end{equation}
Next consider the probability that the entropy production is $X$ 
during the time interval $\tau$. 
We denote this probability by $p_{\tau}^{\mbox{\scriptsize F}}(X)$.  
$p_{\tau}^{\mbox{\scriptsize F}}(X)$ is given by 
integrating eq.~(\ref{eq:PtauXnugamma}) with respect to $\gamma$. 
Thus, we have
\begin{eqnarray}
p_{\tau}^{\mbox{\scriptsize F}}(X) 
& = & 
\int d\gamma \; p_{\tau}^{\mbox{\scriptsize F}}(X;\gamma) 
\nonumber \\
& = & 
\int d\gamma \int d\Gamma_{s} \; \rho_{\ell}(\Gamma_{s}) 
\delta(({\cal T}(\Gamma))_{s}-\gamma) 
\delta(X-\Sigma[\Gamma_{s}\rightarrow({\cal T}(\Gamma))_{s}]) \nonumber \\
& = & 
\int d\Gamma_{s} \; \rho_{\ell}(\Gamma_{s}) 
\delta(X-\Sigma[\Gamma_{s}\rightarrow({\cal T}(\Gamma))_{s}]) \nonumber \\
& = &
\int d\Gamma_{s} \; \rho_{\ell}(\Gamma_{s}) 
e^{\Sigma[\Gamma_{s}\rightarrow ({\cal T}(\Gamma))_{s}]}
e^{-\Sigma[\Gamma_{s}\rightarrow ({\cal T}(\Gamma))_{s}]}
\delta(X-\Sigma[\Gamma_{s}\rightarrow({\cal T}(\Gamma))_{s}]) .
\label{eq:ptauFX}
\end{eqnarray}
Inserting ${\cal T}{\cal T}^{-1}$ in the above equation, we get 
\begin{eqnarray}
p_{\tau}^{\mbox{\scriptsize F}}(X) 
& = &
\int d\Gamma_{s} \; \rho_{\ell}(\Gamma_{s}) {\cal T}{\cal T}^{-1}
e^{\Sigma[\Gamma_{s}\rightarrow({\cal T}(\Gamma))_{s}]}
e^{-\Sigma[\Gamma_{s}\rightarrow({\cal T}(\Gamma))_{s}]}
\delta(X-\Sigma[\Gamma_{s}\rightarrow({\cal T}(\Gamma))_{s}]) \nonumber \\
& = &
\int d\Gamma_{s} \; \rho_{\ell}(({\cal T}^{-1}(\Gamma))_{s}) 
e^{\Sigma[({\cal T}^{-1}(\Gamma))_{s}\rightarrow \Gamma_{s}]}
e^{-\Sigma[({\cal T}^{-1}(\Gamma))_{s}\rightarrow \Gamma_{s}]}
\delta(X+\Sigma[({\cal T}^{-1}(\Gamma))_{s}\rightarrow \Gamma_{s}]).
\end{eqnarray}
Here we have used 
${\cal T}^{-1}(\Sigma[\Gamma_{s}\rightarrow ({\cal T}(\Gamma))_{s}])=
-\Sigma[({\cal T}^{-1}(\Gamma))_{s}\rightarrow \Gamma_{s}]$.
In the last line, we have used the result of the Appendix B 
(i.e., anti-linearity). 
\begin{equation}
\int d\Gamma \; F(\Gamma) {\cal T} G(\Gamma) = 
\int d\Gamma \; G(\Gamma) F({\cal T}^{-1}(\Gamma)).
\end{equation}
Taking $\lim_{\tau\rightarrow\infty} \int dX\; e^{-\lambda X}$ 
from the left hand side, we get
\begin{eqnarray}
\lim_{\tau\rightarrow\infty} \int dX\; e^{-\lambda X} p_{\tau}^{\mbox{\scriptsize F}}(X) 
& = & 
\lim_{\tau\rightarrow\infty} \int dX \; e^{-\lambda X} 
\int d\Gamma_{s} \; \rho_{\ell}(({\cal T}^{-1}(\Gamma))_{s})
\nonumber \\
& & 
\times 
e^{\Sigma[({\cal T}^{-1}(\Gamma))_{s}\rightarrow \Gamma_{s}]}
e^{-\Sigma[({\cal T}^{-1}(\Gamma))_{s}\rightarrow \Gamma_{s}]}
\delta(X+\Sigma[({\cal T}^{-1}(\Gamma))_{s}\rightarrow \Gamma_{s}]) 
\nonumber \\
& = & 
\lim_{\tau\rightarrow\infty} 
\int d\Gamma_{s} \; \rho_{\ell}(({\cal T}^{-1}(\Gamma))_{s}) 
e^{\Sigma[({\cal T}^{-1}(\Gamma))_{s}\rightarrow \Gamma_{s}]}
e^{-\Sigma[({\cal T}^{-1}(\Gamma))_{s}\rightarrow \Gamma_{s}]} 
e^{\lambda \Sigma[({\cal T}^{-1}(\Gamma))_{s}\rightarrow \Gamma_{s}]} \nonumber \\
& = &
\lim_{\tau\rightarrow\infty} Z
\int d\Gamma_{s} \; \rho_{\mbox{\scriptsize ss}}(\Gamma_{s}) 
e^{-(1-\lambda) \Sigma[({\cal T}^{-1}(\Gamma))_{s}\rightarrow \Gamma_{s}]} .
\end{eqnarray} 
In the last line, we have used eq.~(\ref{eq:MZdist_derived}). 
Thus, we obtain
\begin{eqnarray}
\lim_{\tau\rightarrow\infty} \int dX\; e^{-\lambda X} p_{\tau}^{\mbox{\scriptsize F}}(X)
& = & 
\lim_{\tau\rightarrow \infty} Z
\left \langle 
\exp \left [
-(1-\lambda) \sum_{m} \int d{\bf x} \int_{-\tau}^{0} dt\; 
{\bf j}_{m}(\Gamma_{s},{\bf x};t)\cdot {\bf X}_{m}({\bf x}) 
\right ]
\right \rangle_{\mbox{\scriptsize ss}} \nonumber \\
& = & 
\lim_{\tau\rightarrow \infty} Z
\left \langle 
\exp \left [
-(1-\lambda) \sum_{m}\int d{\bf x} \int_{0}^{\tau} dt\; 
{\bf j}_{m}(\Gamma_{s},{\bf x};t)\cdot {\bf X}_{m}({\bf x})
\right ]
\right \rangle_{\mbox{\scriptsize ss}}.
\label{eq:lhs_ft_entropy}
\end{eqnarray}
In the last line, we have used the stationarity of the process, 
namely we can shift the integral domain in the time integral. 

On the other hand, 
going back to eq. (\ref{eq:ptauFX}), and 
using the nonequilibrium detailed balance relation, 
eq.~(\ref{eq:NDBR}), 
we get 
\begin{eqnarray}
p_{\tau}^{\mbox{\scriptsize F}}(X) 
& = & 
\int d\gamma \int d\Gamma_{s} \; 
P_{\nu}(\Gamma_{s}) \delta(({\cal T}(\Gamma))_{s}-\gamma)
\delta(X-\Sigma[\Gamma_{s} \rightarrow ({\cal T}(\Gamma))_{s}]) \nonumber \\
& = & 
\int d\gamma \int d\Gamma_{s} \; 
P_{\nu}(\Gamma_{s}) \delta({\cal T}({\cal T}(\Gamma)^{*})_{s}-\nu^{*}) 
e^{\Sigma[\Gamma_{s}\rightarrow ({\cal T}(\Gamma))_{s}]}
\delta(X-\Sigma[\Gamma_{s}\rightarrow ({\cal T}(\Gamma))_{s}]) \nonumber \\
& = & 
\int d\gamma \int d\Gamma_{s} \; 
\rho_{\ell}(\Gamma_{s}) \delta({\cal T}({\cal T}(\Gamma)^{*})_{s}-\nu^{*}) 
e^{\Sigma[\Gamma_{s} \rightarrow ({\cal T}(\Gamma))_{s}]}
\delta(X-\Sigma[\Gamma_{s} \rightarrow ({\cal T}(\Gamma))_{s}]).
\label{eq:FT1}
\end{eqnarray}
Using the relation ${\cal T}({\cal T}(\Gamma)^{*})= 
{\cal T}(({\cal T}^{-1}(\Gamma^{*})))=\Gamma^{*}$, 
we have 
\begin{eqnarray}
p_{\tau}^{\mbox{\scriptsize F}}(X) 
& = & 
\int d\gamma \int d\Gamma_{s} \;
\rho_{\ell}(\Gamma_{s}) \delta(\Gamma_{s}^{*}-\nu^{*}) 
e^{\Sigma[\Gamma_{s} \rightarrow ({\cal T}(\Gamma))_{s}]}
\delta(X-\Sigma[\Gamma_{s} \rightarrow ({\cal T}(\Gamma))_{s}]) \nonumber \\
& = & 
\int d\gamma \int d\Gamma_{s} \; 
\rho_{\ell}(\Gamma_{s}) \delta(\Gamma_{s}-\nu) 
e^{\Sigma[\Gamma_{s} \rightarrow ({\cal T}(\Gamma))_{s}]}
\delta(X-\Sigma[\Gamma_{s} \rightarrow ({\cal T}(\Gamma))_{s}]) \nonumber \\
& = & 
\int d\gamma \; 
\rho_{\ell}(\nu)
e^{\Sigma[\nu \rightarrow \gamma]}
\delta(X-\Sigma[ \nu \rightarrow (\Gamma ')_{s} ]).
\end{eqnarray}
We take $\lim_{\tau\rightarrow \infty}\int dX\;e^{-\lambda X}$ 
from the left hand side and note that $\gamma=(\Gamma ')_{s}= 
({\cal T}(\Gamma))_{s}$. Then we get
\begin{eqnarray}
\lim_{\tau\rightarrow \infty}
\int dX\;e^{-\lambda X} p_{\tau}^{\mbox{\scriptsize F}}(X)
& = & 
\lim_{\tau\rightarrow \infty} 
\int dX\; e^{-\lambda X} \int d\gamma \; \rho_{\ell}(\nu) 
e^{\Sigma[\nu \rightarrow \gamma]} 
\delta(X - \Sigma[\nu \rightarrow \gamma]) 
\nonumber \\
& = & 
\lim_{\tau\rightarrow \infty} 
\int d\gamma \; \rho_{\ell}(\nu) 
e^{\Sigma[\nu \rightarrow \gamma]} 
e^{-\lambda\Sigma[\nu \rightarrow \gamma]} \nonumber \\
& = & 
\lim_{\tau\rightarrow \infty} Z
\int d\gamma \; \rho_{\mbox{\scriptsize ss}}(\gamma) 
e^{-\lambda\Sigma[\nu \rightarrow \gamma]}.
\end{eqnarray}
In the last line, we used eq.~(\ref{eq:MZdist_derived}). 
Thus, we have
\begin{eqnarray}
\lim_{\tau\rightarrow \infty}
\int dX\;e^{-\lambda X} p_{\tau}^{\mbox{\scriptsize F}}(X)
&= & 
\lim_{\tau\rightarrow \infty} Z
\left \langle
\exp \left [
-\lambda \sum_{m} \int d{\bf x} \int_{-\tau}^{0} dt'\; 
{\bf j}_{m}(\Gamma_{s},{\bf x};t')\cdot {\bf X}_{m}({\bf x}) 
\right ]
\right \rangle_{\mbox{\scriptsize ss}} \nonumber \\
&= & 
\lim_{\tau\rightarrow \infty} Z
\left \langle
\exp \left [
-\lambda \sum_{m} \int d{\bf x} \int_{0}^{\tau} dt'\; 
{\bf j}_{m}(\Gamma_{s},{\bf x};t')\cdot {\bf X}_{m}({\bf x}) 
\right ]
\right \rangle_{\mbox{\scriptsize ss}}.
\label{eq:rhs_ft_entropy}
\end{eqnarray}
In the last line, we have used the stationarity of the process. 
Equations (\ref{eq:lhs_ft_entropy}) and (\ref{eq:rhs_ft_entropy}) imply 
the Lebowitz-Spohn symmetry $\mu(\lambda) = \mu(1-\lambda)$, 
which is the desired result. 
\section{Fluctuation theorem for the current \label{sec5}} 
In this section, we prove the fluctuation theorem for the currents 
for Hamiltonian dynamical systems. 
For the master equation, the fluctuation theorem for the currents 
was proved by Andrieux and Gaspard in ref.~\cite{AndrieuxGaspard07b}. 

We define the cumulant generating functional. 
\begin{equation}
Q( \mbox{\boldmath $\lambda$};{\bf X}) = 
\lim_{\tau\rightarrow \infty} - \frac{1}{\tau}\ln
\left \langle
\exp \left [
-\sum_{m} \int d{\bf x} \int_{0}^{\tau} dt\; 
{\bf j}_{m}(\Gamma_{s},{\bf x};t)\cdot 
\mbox{\boldmath $\lambda$}_{m}({\bf x})
\right]
\right \rangle_{\mbox{\scriptsize ss}}.
\label{eq:QlA}
\end{equation}
Note that the dependence on the thermodynamic forces ${\bf X}({\bf x})$ 
comes from the McLennan-Zubarev steady state distribution. 
We may use the following notation.
\begin{equation}
\mbox{\boldmath $\lambda$}({\bf x}) = 
\left (
\begin{array}{c}
\mbox{\boldmath $\lambda$}_{0}({\bf x})\\
\mbox{\boldmath $\lambda$}_{1}({\bf x})\\
\mbox{\boldmath $\lambda$}_{2}({\bf x})
\end{array}
\right ).
\end{equation}
Our target is 
\begin{equation}
Q( \mbox{\boldmath $\lambda$};{\bf X}) = 
Q( {\bf X}-\mbox{\boldmath $\lambda$};{\bf X}),  
\label{eq:FT_Q}
\end{equation}
which is the fluctuation theorem for the currents. 
We call this relation the Andrieux-Gaspard symmetry. 

First, we define the following quantity. 
\begin{equation}
\Xi[\nu\rightarrow\gamma] 
= \sum_{m} \int d{\bf x} \int_{-\tau}^{0} dt\; 
{\bf j}_{m}(\Gamma,{\bf x};t)\cdot 
\mbox{\boldmath $\lambda$}_{m}({\bf x}). 
\end{equation}
We are interested in how the system has the distribution of $\Xi$ 
in time evolution. 
Thus, we define the probability that $\Xi$ equals $Y$ along a path 
$\nu\rightarrow \gamma$ during the time interval $\tau$ 
by $P_{\tau}^{\mbox{\scriptsize F}}(Y;\gamma)$. 
$P_{\tau}^{\mbox{\scriptsize F}}(Y;\gamma)$ is given by 
\begin{eqnarray}
P_{\tau}^{\mbox{\scriptsize F}}(Y;\gamma) 
& = &
\int d\Gamma_{s} \; P_{\nu}(\Gamma_{s}) \delta(({\cal T}(\Gamma))_{s}-\gamma) 
\delta(Y-\Xi[\Gamma_{s}\rightarrow ({\cal T}(\Gamma))_{s}]) \nonumber \\
& = &
\int d\Gamma_{s} \; 
\rho_{\ell}(\Gamma_{s})
\delta(({\cal T}(\Gamma))_{s}-\gamma) 
\delta(Y-\Xi[\Gamma_{s}\rightarrow ({\cal T}(\Gamma))_{s}]) .
\label{eq:PtauYnugamma}
\end{eqnarray}
Next, we consider the probability that $\Xi$ is $Y$ during the time 
interval $\tau$. 
We denote this probability by $P_{\tau}^{\mbox{\scriptsize F}} (Y)$. 
$P_{\tau}^{\mbox{\scriptsize F}} (Y)$ is given by integrating 
eq.~(\ref{eq:PtauYnugamma}) with respect to $\gamma$. 
Thus, we have 
\begin{eqnarray}
P_{\tau}^{\mbox{\scriptsize F}} (Y) 
& = &
\int d\gamma \int d\Gamma_{s} \; P_{\tau}^{\mbox{\scriptsize F}}(Y;\gamma) 
\nonumber \\
& = &
\int d\gamma \int d\Gamma_{s} \; \rho_{\ell}(\Gamma_{s})
\delta(({\cal T}(\Gamma))_{s}-\gamma) 
\delta(Y-\Xi[\Gamma_{s}\rightarrow ({\cal T}(\Gamma))_{s}]) 
\nonumber \\
& = &
\int d\Gamma_{s} \; \rho_{\ell}(\Gamma_{s})
\delta(Y-\Xi[\Gamma_{s}\rightarrow ({\cal T}(\Gamma))_{s}]) 
\nonumber \\
& = &
\int d\Gamma_{s} \; \rho_{\ell}(\Gamma_{s})
e^{-\Sigma[\Gamma_{s}\rightarrow ({\cal T}(\Gamma))_{s}]}
e^{\Sigma[\Gamma_{s}\rightarrow ({\cal T}(\Gamma))_{s}]}
\delta(Y-\Xi[\Gamma_{s}\rightarrow ({\cal T}(\Gamma))_{s}]) .
\label{eq:ptauFY}
\end{eqnarray}
Inserting ${\cal T}{\cal T}^{-1}$ in the above equation, we have 
\begin{eqnarray}
P_{\tau}^{\mbox{\scriptsize F}} (Y) 
& = &
\int d\Gamma_{s} \; 
\rho_{\ell}(\Gamma_{s}) {\cal T}{\cal T}^{-1}
e^{-\Sigma[\Gamma_{s}\rightarrow ({\cal T}(\Gamma))_{s} ]}
e^{\Sigma[\Gamma_{s}\rightarrow ({\cal T}(\Gamma))_{s} ]}
\delta(Y-\Xi[\Gamma_{s} \rightarrow ({\cal T}(\Gamma))_{s} ]) \nonumber \\
& = & 
\int d\Gamma_{s} \; 
\rho_{\ell}(({\cal T}^{-1}(\Gamma))_{s}) 
e^{-\Sigma[({\cal T}^{-1}(\Gamma))_{s}\rightarrow \Gamma_{s}]}
e^{\Sigma[({\cal T}^{-1}(\Gamma))_{s}\rightarrow \Gamma_{s}]}
\delta(Y+\Xi[({\cal T}^{-1}(\Gamma))_{s}\rightarrow \Gamma_{s}]).
\end{eqnarray}
In the last line, we used the result of the Appendix B (i.e., anti-linearity). 
Taking the limit and integration of
$\lim_{\tau\rightarrow \infty} \int dY\; e^{-Y}$, 
we obtain
\begin{eqnarray}
\lim_{\tau\rightarrow\infty} \int dY\; e^{-Y} P_{\tau}^{\mbox{\scriptsize F}}(Y) 
& = & 
\lim_{\tau\rightarrow\infty} \int dY\; e^{-Y} 
\int d\Gamma_{s} \; 
\rho_{\ell}(({\cal T}^{-1}(\Gamma))_{s}) 
e^{\Sigma[({\cal T}^{-1}(\Gamma))_{s}\rightarrow \Gamma_{s}]}
e^{-\Sigma[({\cal T}^{-1}(\Gamma))_{s}\rightarrow \Gamma_{s}]}
\delta(Y+\Xi[({\cal T}^{-1}(\Gamma))_{s}\rightarrow \Gamma_{s}]) \nonumber \\
& = & 
\lim_{\tau\rightarrow\infty} 
\int d\Gamma_{s} \; 
\rho_{\ell}(({\cal T}^{-1}(\Gamma))_{s}) 
e^{\Sigma[({\cal T}^{-1}(\Gamma))_{s}\rightarrow \Gamma_{s}]}
e^{-\Sigma[({\cal T}^{-1}(\Gamma))_{s}\rightarrow \Gamma_{s}]}
e^{\Xi[({\cal T}^{-1}(\Gamma))_{s}\rightarrow \Gamma_{s}]} \nonumber \\
& = & 
\lim_{\tau\rightarrow\infty} 
Z \int d\Gamma_{s} \; 
\rho_{\mbox{\scriptsize ss}}(\Gamma_{s}) 
e^{-\Sigma[({\cal T}^{-1}(\Gamma))_{s}\rightarrow \Gamma_{s}]}
e^{\Xi[({\cal T}^{-1}(\Gamma))_{s}\rightarrow \Gamma_{s}]}. 
\end{eqnarray}
Thus, we finally have 
\begin{eqnarray}
\lim_{\tau\rightarrow\infty} \int dY\; e^{-Y} P_{\tau}^{\mbox{\scriptsize F}}(Y)
& = & 
\lim_{\tau\rightarrow\infty} 
Z \int d\Gamma_{s} \; 
\rho_{\mbox{\scriptsize ss}}(\Gamma_{s}) 
e^{-\Sigma[({\cal T}^{-1}(\Gamma))_{s}\rightarrow \Gamma_{s}]}
e^{\Xi[({\cal T}^{-1}(\Gamma))_{s}\rightarrow \Gamma_{s}]} \nonumber\\
& = &
\lim_{\tau\rightarrow\infty} Z
\left \langle
\exp \left [
-\sum_{m} \int d{\bf x} \int_{-\tau}^{0} dt \; {\bf j}_{m}(\Gamma_{s},{\bf x};t)
\cdot ({\bf X}_{m}({\bf x})-\mbox{\boldmath $\lambda$}_{m}({\bf x}))
\right ]
\right \rangle_{\mbox{\scriptsize ss}} \nonumber \\
& = &
\lim_{\tau\rightarrow\infty} Z
\left \langle
\exp \left [
-\sum_{m} \int d{\bf x} \int_{0}^{\tau} dt \; {\bf j}_{m}(\Gamma_{s},{\bf x};t)
\cdot ({\bf X}_{m}({\bf x})-\mbox{\boldmath $\lambda$}_{m}({\bf x}))
\right ]
\right \rangle_{\mbox{\scriptsize ss}}.
\label{eq:l}
\end{eqnarray}
In the second line, we have used eq.~(\ref{eq:MZdist_derived}). 
In the last line, we have used the stationarity of the process.

On the other hand, if we use the nonequilibrium detailed balance relation, 
eq.~(\ref{eq:NDBR}) to eq.(\ref{eq:ptauFY}), we have
\begin{eqnarray}
P_{\tau}^{\mbox{\scriptsize F}}(Y) 
& = & 
\int d\gamma \int d\Gamma_{s}\; 
P_{\nu}(\Gamma_{s}) \delta(({\cal T}(\Gamma))_{s}-\gamma) 
\delta(Y-\Xi[\Gamma_{s} \rightarrow ({\cal T}(\Gamma))_{s}]) \nonumber \\
& = & 
\int d\gamma \int d\Gamma_{s}\; 
P_{\nu}(\Gamma_{s}) 
\delta(({\cal T}({\cal T}(\Gamma))^{*})_{s}-\nu^{*}) 
e^{\Sigma[\Gamma_{s}\rightarrow ({\cal T}(\Gamma))_{s}]}
\delta(Y-\Xi[\Gamma_{s} \rightarrow ({\cal T}(\Gamma))_{s}]) \nonumber \\
& = & 
\int d\gamma \int d\Gamma_{s}\; 
\rho_{\ell}(\Gamma_{s}) 
\delta(({\cal T}({\cal T}(\Gamma))^{*})_{s}-\nu^{*}) 
e^{\Sigma[\Gamma_{s} \rightarrow ({\cal T}(\Gamma))_{s}]}
\delta(Y-\Xi[\Gamma_{s} \rightarrow ({\cal T}(\Gamma))_{s}]) \nonumber \\
& = & 
\int d\gamma \int d\Gamma_{s}\; 
\rho_{\ell}(\Gamma_{s}) 
\delta(\Gamma_{s}^{*}-\nu^{*}) 
e^{\Sigma[\Gamma_{s} \rightarrow ({\cal T}(\Gamma))_{s}]}
\delta(Y-\Xi[\Gamma_{s} \rightarrow ({\cal T}(\Gamma))_{s}]) \nonumber \\
& = & 
\int d\gamma \; 
\rho_{\ell}(\nu)
e^{\Sigma[\nu \rightarrow ({\cal T}(\Gamma))_{s}]}
\delta(Y-\Xi[\nu \rightarrow ({\cal T}(\Gamma))_{s}]).
\label{eq:FT2}
\end{eqnarray}
Here we used the relation ${\cal T}({\cal T}(\Gamma)^{*})= \Gamma^{*}$.  
Then, taking the limit and integration of
$\lim_{\tau\rightarrow \infty} \int dY\; e^{-Y}$, we have 
\begin{eqnarray}
\lim_{\tau\rightarrow \infty} \int dY\; e^{-Y} P_{\tau}^{\mbox{\scriptsize F}}(Y)
& = & 
\lim_{\tau\rightarrow \infty} \int dY\; e^{-Y} 
\int d\gamma \; 
\rho_{\ell}(\nu)
e^{\Sigma[\nu \rightarrow ({\cal T}(\Gamma))_{s}]}
\delta(Y-\Xi[\nu \rightarrow ({\cal T}(\Gamma))_{s}]) \nonumber \\
& = & 
\lim_{\tau\rightarrow \infty} 
\int d\gamma \; 
\rho_{\ell}(\nu)
e^{\Sigma[\nu \rightarrow ({\cal T}(\Gamma))_{s}]}
e^{-\Xi[\nu \rightarrow ({\cal T}(\Gamma))_{s}]} .
\end{eqnarray}
Here we note that $\gamma=({\cal T}(\Gamma))_{s}$. 
We obtain 
\begin{eqnarray}
\lim_{\tau\rightarrow \infty} \int dY\; e^{-Y} P_{\tau}^{\mbox{\scriptsize F}}(Y)
& = & 
\lim_{\tau\rightarrow \infty} Z
\int d\gamma \; 
\rho_{\mbox{\scriptsize ss}}(\gamma)
e^{-\Xi[\nu \rightarrow \gamma]} \nonumber \\
& = & 
\lim_{\tau \rightarrow \infty} Z
\left \langle
\exp \left [
- \sum_{m}\int d{\bf x} \int_{-\tau}^{0} dt\; 
{\bf j}_{m}(\Gamma_{s},{\bf x};t)\cdot 
\mbox{\boldmath $\lambda$}_{m}({\bf x}) 
\right ]
\right \rangle_{\mbox{\scriptsize ss}} \nonumber \\
& = & 
\lim_{\tau\rightarrow \infty} Z
\left \langle
\exp \left [
- \sum_{m} \int d{\bf x} \int_{0}^{\tau} dt\; 
{\bf j}_{m}(\Gamma_{s},{\bf x};t)\cdot 
\mbox{\boldmath $\lambda$}_{m}({\bf x}) 
\right ]
\right \rangle_{\mbox{\scriptsize ss}}.
\label{eq:Al}
\end{eqnarray}
In the first line, we have used eq.~(\ref{eq:MZdist_derived}). 
In the last line, we have used the stationarity of the process. 
Thus, by eqs.~(\ref{eq:l}) and (\ref{eq:Al}), we obtain 
\begin{eqnarray}
& & \lim_{\tau\rightarrow \infty} 
\left \langle
\exp \left [
- \sum_{m}\int d{\bf x} \int_{0}^{\tau} dt'\; 
{\bf j}_{m}(\Gamma_{s},{\bf x};t')\cdot 
\mbox{\boldmath $\lambda$}_{m}({\bf x}) 
\right ]
\right \rangle_{\mbox{\scriptsize ss}} \nonumber \\
& = &
\lim_{\tau\rightarrow \infty} 
\left \langle
\exp \left [
- \sum_{m}\int d{\bf x} \int_{0}^{\tau} dt'\; 
{\bf j}_{m}(\Gamma_{s},{\bf x};t')\cdot 
({\bf X}_{m}({\bf x})-\mbox{\boldmath $\lambda$}_{m}({\bf x}))
\right ]
\right \rangle_{\mbox{\scriptsize ss}}.
\end{eqnarray}
By the definition of the cumulant generating functional, eq.~(\ref{eq:QlA}), 
this gives the fluctuation theorem (i.e., Andrieux-Gaspard symmetry) 
$Q(\mbox{\boldmath $\lambda$};{\bf X}) = 
Q({\bf X}-\mbox{\boldmath $\lambda$};{\bf X})$.
\section{On the nonequilibrium detailed balance relation}
\label{sec6}

Some readers may feel that the assumption of 
the nonequilibrium detailed balance relation is 
a key relation, and that it is "doubtful" whether it is formed. 
In this section, we examine the nonequilibrium detailed balance relation, 
eq.(\ref{eq:NDBR}).  

Back to the derivation of the fluctuation theorem 
for the entropy production, 
remind that the nonequilibrium detailed balance relation is used 
at eq.(\ref{eq:MZ_NEDBR}) 
(the McLennan-Zubarev steady state distribution),  
at eq.(\ref{eq:FT1}) 
(the fluctuation theorem for the entropy production), 
and at eq.(\ref{eq:FT2}) (the fluctuation theorem for the current).
Suppose that the nonequilibrium detailed balance relation 
is given by
\begin{equation}
P[\nu \rightarrow \gamma] 
= e^{\Theta[\nu\rightarrow \gamma]}P[\gamma^{*}\rightarrow \nu^{*}],
\end{equation}
where the function $\Theta[\nu\rightarrow\gamma]$ 
is assumed not to be $\Sigma[\nu\rightarrow \gamma]$.
Then, following the derivation of the McLennan-Zubarev steady state 
distribution,
we obtain the following steady state distribution.
\begin{equation}
\rho_{\overline{ss}}(\gamma) = Z_{*}^{-1} \rho_{\ell}(\nu) 
\exp(\Theta[\nu\rightarrow \gamma]),
\end{equation}
where
\begin{equation}
Z_{*} = \int d\gamma \;  \rho_{\ell}(\nu) \exp(\Theta[\nu\rightarrow \gamma]).
\end{equation}
Further following the derivation of the fluctuation theorem for 
the entropy production, we have
\begin{equation}
\lim_{\tau\rightarrow \infty} Z_{*} 
\left \langle e^{-\lambda\Sigma} \right \rangle_{\overline{ss}} =
\lim_{\tau\rightarrow \infty} Z_{*} 
\left \langle e^{\lambda\Sigma - \Theta} \right \rangle_{\overline{ss}},
\end{equation}
where
\begin{equation}
\left \langle \cdots \right \rangle_{\overline{ss}}
= \int d\gamma \; \rho_{\overline{ss}} (\gamma) \cdots.
\end{equation}
We find that the fluctuation theorem is not valid 
for the general case of $\Theta([\nu\rightarrow \gamma])$. 
It is clear that the fluctuation theorem is valid 
when the condition 
\begin{equation}
\Theta[\nu\rightarrow \gamma] = 
\Sigma[\nu\rightarrow \gamma],
\end{equation}
is formed. 
Thus, assuming the nonequilibrium detailed balance relation, implies 
the validity of the fluctuation theorem for the entropy production.
The same argument for the fluctuation theorem for the current is true. 
Following the derivation of the fluctuation theorem for the current, 
we have
\begin{equation}
\lim_{\tau\rightarrow \infty} Z_{*} 
\left \langle e^{-\Theta+\Xi}\right \rangle_{\overline{ss}}
=
\lim_{\tau\rightarrow \infty} Z_{*} 
\left \langle e^{-\Xi}\right \rangle_{\overline{ss}}.
\end{equation}
The fluctuation theorem for the current is valid 
if and only if the condition 
$\Theta[\nu\rightarrow \gamma] = \Sigma[\nu\rightarrow \gamma]$ consists. 
Thus, assuming the nonequilibrium detailed balance relation 
as eq.(\ref{eq:NDBR}) implies the validity of two fluctuation theorems. 

We may want to justify the nonequilibrium detailed balance relation. 
However, unfortunately, there is no plausible justification. 
We believe the analogy to similar relations derived for 
other types of dynamics, for instance, 
stochastic Markovian dynamics\cite{Crooks98,Crooks99}.
Thus, we assume the nonequilibrium detailed balance relation, 
eq.(\ref{eq:NDBR}). 
As a result, two fluctuation theorems hold good.

\section{Nonlinear response \label{sec7}} 
The cumulant generating functional for the current, 
i.e. eq.~(\ref{eq:QlA}), is useful for evaluating the mean current. 
The current can be written in terms of thermodynamic forces, 
i.e., a power series of thermodynamic forces. 
The thermodynamic force measures the strength of 
the nonequilibrium property, i.e., 
how away from equilibrium. 
In order to clarify the strength of the nonequilibrium property,  
here we set ${\bf X}({\bf x})$ as $\epsilon {\bf X}({\bf x})$. 
The McLennan-Zubarev steady state distribution depends 
on the thermodynamic forces.
We can expand the McLennan-Zubarev steady state distribution 
in a power series of thermodynamic forces. 
Then, the mean current is a sum of 
the equilibrium current ($O(\epsilon^{0})$), 
the linear response ($O(\epsilon^{1})$) (the Green-Kubo formula), 
and the non-linear responses ($O(\epsilon^{n}),\;\; n\geq 2$).
This idea is originally due to Gallavotti\cite{Gallavotti96}. 
Gallavotti's idea was applied to the master equation 
by Andrieux and Gaspard\cite{AndrieuxGaspard04,AndrieuxGaspard07a} 
to give the non-linear response. 
In this section, we evaluate the linear response and non-linear response 
in the current for the case of the McLennan-Zubarev steady distribution.
We set 
${\bf X}={\bf X}'/k_{B}$ and 
$\mbox{\boldmath $\lambda$} = \mbox{\boldmath $\lambda$}'/k_{B}$.
The cumulant generating functional is 
$Q(\mbox{\boldmath $\lambda$};{\bf X})
=\tilde{Q}(\mbox{\boldmath $\lambda$}';{\bf X}')$.

The $\alpha$th component $J_{\alpha}({\bf x})$ of the mean current 
at ${\bf x}$ is given by the functional derivative 
with respect to $\lambda_{\alpha}({\bf x})$ 
\begin{equation}
J_{\alpha}({\bf x}) = 
\left . \frac{\delta Q}{\delta \lambda_{\alpha}({\bf x})} 
\right |_{\mbox{\boldmath $\lambda$}={\bf 0}}
= 
k_{B} 
\left . \frac{\delta \tilde{Q}}{\delta \lambda'_{\alpha}({\bf x})} 
\right |_{\mbox{\boldmath $\lambda$}'={\bf 0}}.
\label{eq:mean_current_1}
\end{equation}
By definition, we have
\begin{equation}
J_{\alpha}({\bf x}) = 
\lim_{t\rightarrow \infty} \frac{1}{t} \int_{0}^{t} dt'\; 
\langle j_{\alpha}(\Gamma,{\bf x};t')\rangle_{\mbox{\scriptsize ss}}.
\end{equation}
The mean current is expanded in thermodynamic forces.
\begin{eqnarray}
J_{\alpha}({\bf x}) 
& = & 
J_{\alpha}^{(\mbox{\scriptsize eq})}({\bf x}) 
+ 
\frac{\epsilon}{1!} \sum_{\beta}
\int d{\bf x}_{1}\; {\cal L}_{\alpha\beta}({\bf x},{\bf x}_{1}) 
X'_{\beta}({\bf x}_{1}) \nonumber \\
& & 
+ \frac{\epsilon^{2}}{2!} 
\sum_{\beta,\gamma}
\int \!\!\! \int d{\bf x}_{1}d{\bf x}_{2}\; 
{\cal M}_{\alpha\beta\gamma}({\bf x},{\bf x}_{1},{\bf x}_{2}) 
X'_{\beta}({\bf x}_{1}) X'_{\gamma}({\bf x}_{2}) \nonumber \\
& & 
+ \frac{\epsilon^{3}}{3!} 
\sum_{\beta,\gamma,\delta}
\int \!\!\! \int \!\!\! \int d{\bf x}_{1}d{\bf x}_{2}d{\bf x}_{3}\; 
{\cal N}_{\alpha\beta\gamma\delta}
({\bf x},{\bf x}_{1},{\bf x}_{2},{\bf x}_{3}) 
X'_{\beta}({\bf x}_{1}) X'_{\gamma}({\bf x}_{2}) X'_{\delta}({\bf x}_{3})
\nonumber \\
& & 
+ \cdots ,
\end{eqnarray}
where
\begin{eqnarray}
J_{\alpha}^{(\mbox{\scriptsize eq})}({\bf x}) & = & 
\lim_{t\rightarrow \infty} \frac{1}{t} \int_{0}^{t} dt'\; 
\langle j_{\alpha}({\bf x};t')\rangle_{\mbox{\scriptsize eq}} \nonumber \\
& = & 
\int d\Gamma \; \rho_{\mbox{\scriptsize can}}(\Gamma) j_{\alpha}(\Gamma,{\bf x};0) 
\nonumber \\
& = & \langle j_{\alpha}(\Gamma,{\bf x};0) \rangle_{\mbox{\scriptsize eq}}. 
\end{eqnarray}
$\langle \cdots \rangle_{\mbox{\scriptsize eq}}$ is the average 
over the canonical ensemble. 
Here we consider that the system in local equilibrium is near equilibrium, 
that is, the temperature field is almost constant and ${\bf u}\approx 0$. 
So $\rho_{\ell}$ is replaced by $\rho_{\mbox{\scriptsize can}}$.
We have used the fact that 
the time-dependence is washed out in the average over the canonical ensemble, 
i.e., the average over the equilibrium state. 

On the other hand, we obtain the mean current by averaging over 
the McLennan-Zubarev steady state distribution.
\begin{equation}
J_{\alpha}({\bf x}) = 
\langle j_{\alpha}(\Gamma,{\bf x};0) \rangle_{\mbox{\scriptsize ss}}
= 
\int d\Gamma \; \rho_{\mbox{\scriptsize ss}}(\Gamma) j_{\alpha}(\Gamma,{\bf x};0). 
\label{eq:mean_current_2}
\end{equation}
Expanding $\rho_{\mbox{\scriptsize ss}}(\Gamma)$ in terms of thermodynamic forces, 
we also obtain the mean current in a power series of thermodynamic forces.   
Of course, each term in two expansions from eqs.~(\ref{eq:mean_current_1}) and 
(\ref{eq:mean_current_2}) should coincide identically.
\subsection{Linear response}
The function ${\cal L}_{\alpha\beta}({\bf x},{\bf x}_{1})$ is given by 
\begin{equation}
{\cal L}_{\alpha\beta}({\bf x},{\bf x}_{1}) 
= 
k_{B}
\frac{\delta^{2}\tilde{Q}}
{\delta \lambda'_{\alpha}({\bf x}) \delta X'_{\beta}({\bf x}_{1})}
({\bf 0};{\bf 0})
=
\frac{1}{k_{B}}
\frac{\delta^{2}Q}
{\delta \lambda_{\alpha}({\bf x}) \delta X_{\beta}({\bf x}_{1})}
({\bf 0};{\bf 0}).
\end{equation}
Differentiating eq.~(\ref{eq:FT_Q}), we obtain
\begin{eqnarray}
\frac{\delta^{2} Q}
{\delta \lambda_{\alpha}({\bf x}) \delta X_{\beta}({\bf x}_{1})}
(\mbox{\boldmath $\lambda$};{\bf X})
& = & 
-\frac{\delta^{2}Q}
{\delta \lambda_{\alpha}({\bf x}) \delta \lambda_{\beta}({\bf x}_{1})}
({\bf X}-\mbox{\boldmath $\lambda$};{\bf X})
\nonumber \\
& & 
- 
\frac{\delta^{2}Q}
{\delta \lambda_{\alpha}({\bf x}) \delta X_{\beta}({\bf x}_{1})}
({\bf X}-\mbox{\boldmath $\lambda$};{\bf X}).
\end{eqnarray}
Therefore, setting $\mbox{\boldmath $\lambda$}={\bf 0}$ 
and ${\bf X}={\bf 0}$, we have 
\begin{equation}
{\cal L}_{\alpha,\beta}({\bf x},{\bf x}_{1})
= 
\frac{1}{k_{B}}
\frac{\delta^{2}Q}
{\delta \lambda_{\alpha}({\bf x}) \delta X_{\beta}({\bf x}_{1})}
({\bf 0};{\bf 0}) 
= 
- \frac{1}{2k_{B}}
\frac{\delta^{2}Q}
{\delta \lambda_{\alpha}({\bf x}) \delta \lambda_{\beta}({\bf x}_{1})}
({\bf 0};{\bf 0}). 
\end{equation}
This gives the Onsager reciprocity relation\cite{Onsager1931a,Onsager1931b}
\begin{equation}
{\cal L}_{\alpha\beta}({\bf x},{\bf x}_{1}) = 
{\cal L}_{\beta\alpha}({\bf x}_{1},{\bf x}). 
\end{equation}
Expanding the McLennan-Zubarev steady state distribution in thermodynamic 
forces, namely expanding the exponential function 
and the normalization constant $Z$, 
we get 
\begin{equation}
{\cal L}_{\alpha\beta}({\bf x},{\bf x}_{1}) 
= \frac{1}{2k_{B}} 
\int_{-\infty}^{\infty} dt\; 
\phi_{\alpha\beta}^{(2)}({\bf x},{\bf x}_{1};t),
\label{eq:Green-Kubo-formula}
\end{equation}
where $\phi_{\alpha\beta}^{(2)}({\bf x},{\bf x}_{1};t)$ is the two-point 
current correlation function, 
\begin{equation}
\phi_{\alpha\beta}^{(2)}({\bf x},{\bf x}_{1};t)
= 
\langle
[j_{\alpha}(\Gamma,{\bf x};t) - 
\langle j_{\alpha}(\Gamma,{\bf x};t)\rangle_{\mbox{\scriptsize eq}}]
[j_{\beta}(\Gamma,{\bf x}_{1};0)-\langle j_{\beta}(\Gamma,{\bf x}_{1};0)
\rangle_{\mbox{\scriptsize eq}}]
\rangle_{\mbox{\scriptsize eq}}.
\end{equation}
Here note that the average is taken over the local equilibrium state. 

Comparing the term in eq.~(\ref{eq:mean_current_1}) 
with that of eq.~(\ref{eq:mean_current_2}) in the order of $O(\epsilon^{1})$, 

\begin{eqnarray}
\int_{0}^{\infty} dt\; 
\phi_{\alpha\beta}^{(2)}({\bf x},{\bf x}_{1};t)
& = & 
\int_{0}^{\infty} dt\; 
\phi_{\alpha\beta}^{(2)}({\bf x},{\bf x}_{1};-t).
\label{eq:nontrivial_2}
\end{eqnarray}
By eq.~(\ref{eq:nontrivial_2}), 
${\cal L}_{\alpha\beta}({\bf x},{\bf x}_{1})$ is given by 
\begin{equation}
{\cal L}_{\alpha\beta}({\bf x},{\bf x}_{1}) 
= 
\frac{1}{k_{B}}
\int_{0}^{\infty} dt\; 
\phi_{\alpha\beta}^{(2)}({\bf x},{\bf x}_{1};t).
\label{eq:Green-Kubo-formula2}
\end{equation}
Equation (\ref{eq:Green-Kubo-formula2}) is nothing but the Green-Kubo formula
\cite{Green1952,Green1954,Kubo1957}. 
\subsection{Non-linear response of $O(\epsilon^{2})$}
The function ${\cal M}_{\alpha\beta\gamma}({\bf x},{\bf x}_{1},{\bf x}_{2})$ 
is given by
\begin{equation}
{\cal M}_{\alpha\beta\gamma}({\bf x},{\bf x}_{1},{\bf x}_{2}) = 
\frac{1}{k_{B}^{2}}
\frac{\delta^{3}Q}{\delta \lambda_{\alpha}({\bf x}) 
\delta X_{\beta}({\bf x}_{1})\delta X_{\gamma}({\bf x}_{2})}({\bf 0};{\bf 0}).
\end{equation}
In the same way of the previous subsection, 
differentiating eq.~(\ref{eq:FT_Q}), 
and setting $\mbox{\boldmath $\lambda$}$ and ${\bf X}=0$,
we obtain
\begin{eqnarray}
{\cal M}_{\alpha\beta\gamma}({\bf x},{\bf x}_{1},{\bf x}_{2})
& = & 
-\frac{1}{2k_{B}^{2}}
\frac{\delta^{3}Q}
{\delta \lambda_{\alpha}({\bf x})\delta \lambda_{\beta}({\bf x}_{1}) 
\delta \lambda_{\gamma}({\bf x}_{2})}
({\bf 0};{\bf 0})
\nonumber \\
& & 
- \frac{1}{2k_{B}^{2}}
\frac{\delta^{3}Q}
{\delta \lambda_{\alpha}({\bf x})\delta \lambda_{\beta}({\bf x}_{1}) 
\delta X_{\gamma}({\bf x}_{2})}
({\bf 0};{\bf 0})
\nonumber \\
& & 
- \frac{1}{2k_{B}^{2}}
\frac{\delta^{3}Q}
{\delta \lambda_{\alpha}({\bf x})\delta \lambda_{\gamma}({\bf x}_{1}) 
\delta X_{\beta}({\bf x}_{2})}
({\bf 0};{\bf 0}).
\label{eq:M_0}
\end{eqnarray}
Evaluating the right hand side of eq.~(\ref{eq:M_0}), it leads 
\begin{eqnarray}
& & 
{\cal M}_{\alpha\beta\gamma}({\bf x},{\bf x}_{1},{\bf x}_{2})
\nonumber \\
& = &
\frac{1}{2k_{B}^{2}}\left \{
\int_{-\infty}^{\infty}dt_{1} \int_{-\infty}^{0}dt_{2} 
+
\int_{-\infty}^{0}dt_{1} \int_{-\infty}^{\infty}dt_{2} 
\right \}
\phi_{\alpha\beta\gamma}^{(3)}({\bf x},{\bf x}_{1},{\bf x}_{2};t_{1},t_{2}),
\label{eq:cal_M}
\end{eqnarray}
where
\begin{eqnarray}
\phi_{\alpha\beta\gamma}^{(3)}({\bf x},{\bf x}_{1},{\bf x}_{2};t_{1},t_{2}) 
& = & 
\langle j_{\alpha}(\Gamma,{\bf x};0)j_{\beta}(\Gamma,{\bf x}_{1};t_{1})
j_{\gamma}(\Gamma,{\bf x}_{2};t_{2}) 
\rangle_{\mbox{\scriptsize eq}}
- 
\langle j_{\alpha}(\Gamma,{\bf x};0) 
j_{\beta}(\Gamma,{\bf x}_{1};t_{1})\rangle_{\mbox{\scriptsize eq}}
\langle j_{\gamma}(\Gamma,{\bf x}_{2};t_{2}) \rangle_{\mbox{\scriptsize eq}}
\nonumber \\
& &
- 
\langle j_{\alpha}(\Gamma,{\bf x};0)\rangle_{\mbox{\scriptsize eq}}
\langle j_{\beta}(\Gamma,{\bf x}_{1};t_{1})
j_{\gamma}(\Gamma,{\bf x}_{2};t_{2}) 
\rangle_{\mbox{\scriptsize eq}}
- 
\langle j_{\alpha}(\Gamma,{\bf x};0) j_{\gamma}(\Gamma,{\bf x}_{2};t_{2}) 
\rangle_{\mbox{\scriptsize eq}}
\langle j_{\beta}(\Gamma,{\bf x}_{1};t_{1}) 
\rangle_{\mbox{\scriptsize eq}}
\nonumber \\
& & 
+2 
\langle j_{\alpha}(\Gamma,{\bf x};0) \rangle_{\mbox{\scriptsize eq}}
\langle j_{\beta}(\Gamma,{\bf x}_{1};t_{1})\rangle_{\mbox{\scriptsize eq}}
\langle j_{\gamma}(\Gamma,{\bf x}_{2};t_{2})\rangle_{\mbox{\scriptsize eq}}.
\end{eqnarray}
On the other hand, we obtain the term in $O(\epsilon^{2})$ 
from eq.~(\ref{eq:mean_current_2}). 
This should coincide with eq.~(\ref{eq:cal_M}). 
This comparison gives a non-trivial relation of 
the three-point current correlation function
\begin{equation}
\left \{ 
\int_{0}^{\infty}dt_{1} \int_{-\infty}^{0}dt_{2}
+ 
\int_{-\infty}^{0}dt_{1} \int_{0}^{\infty}dt_{2}
\right \} 
\phi_{\alpha\beta\gamma}^{(3)}({\bf x},{\bf x}_{1},{\bf x}_{2};t_{1},t_{2})
= 0.
\label{eq:nontrivial_3}
\end{equation}
With this relation, 
${\cal M}_{\alpha\beta\gamma}({\bf x},{\bf x}_{1},{\bf x}_{2})$ is given by 
\begin{eqnarray}
{\cal M}_{\alpha\beta\gamma}({\bf x},{\bf x}_{1},{\bf x}_{2}) 
& = &
\frac{1}{k_{B}^{2}}
\int_{-\infty}^{0}dt_{1} \int_{-\infty}^{0}dt_{2} \; 
\phi_{\alpha\beta\gamma}^{(3)}({\bf x},{\bf x}_{1},{\bf x}_{2};t_{1},t_{2}).
\label{eq:cal_M_2}
\end{eqnarray}
The non-trivial relations of eqs.~(\ref{eq:nontrivial_2}) and 
(\ref{eq:nontrivial_3}) are the consequence of the time-reversal symmetry, 
i.e., the fluctuation theorem 
$Q(\mbox{\boldmath $\lambda$};{\bf X})=
Q({\bf X}-\mbox{\boldmath $\lambda$};{\bf X})$.
In the same way as shown above, non-trivial relations would be obtained 
for the order of $O(\epsilon^{n})\;\; (n\geq 3)$. 
\section{Concluding remarks \label{sec8}} 
In this paper, we have derived the McLennan-Zubarev steady state distribution 
from the nonequilibrium detailed balance relation, 
and have proved the fluctuation theorems for the entropy production 
and for the currents. 
As shown in the derivation and the proof, 
the McLennan-Zubarev steady state distribution 
and the fluctuation theorems have the same root, 
in the sense that they are derived from the same relations, 
i.e., the time-reversal symmetry and 
the nonequilibrium detailed balance relation. 

As a consequence of the fluctuation theorem for the currents, 
we have derived the expression for the mean current 
in terms of thermodynamic forces.
The linear response in the order $O(\epsilon^{1})$ and 
non-linear response in the order of $O(\epsilon^{2})$ are derived.
The linear response exactly coincides with the Green-Kubo formula. 
The non-linear response of $O(\epsilon^{2}$) gives the first correction 
to the Green-Kubo formula. 
We have also obtained the non-trivial relations 
of the correlation functions for the orders of $O(\epsilon^{1})$ and 
$O(\epsilon^{2})$. 
These non-trivial relations of the correlation functions 
are a consequence of time-reversal symmetry. 
The corrections $O(\epsilon^{n}), n \geq 3$ can be derived in a systematic way
\cite{AndrieuxGaspard07a}.
In addition, the results of this paper are exact without approximations 
except replacing the local equilibrium state by the canonical distribution 
for the section of the nonlinear response.


%
%

\section*{Acknowledgements}
The author thanks Eiji Konishi for careful reading the manuscript. 
This work is supported by JSPS KAKENHI (No.24654122).


\renewcommand{\thesection}{\Alph{section}}
\renewcommand{\theequation}{\thesection . \arabic{equation}}
\setcounter{section}{1}
\setcounter{equation}{0}

\section*{Appendix A \label{AppendixA}}

In this appendix, we show McLennan's derivation of 
the McLennan-Zubarev steady state distribution\cite{McLennan59}.
We consider the system which is coupled with baths. 
Now the Hamiltonian is given by
\begin{equation}
H_{\mbox{\scriptsize tot}} = H_{S}+H_{B}+U,
\label{eq:HamiltonianTot}
\end{equation}
where $H_{S}$ is the Hamiltonian of the system, 
$H_{B}$ is the Hamiltonian of the baths, 
and $U$ is the interaction between the system and baths.
$U$ does not include the momenta. 
The whole system is a Hamiltonian system.
The Liouville equation is 
\begin{equation}
\frac{\partial f}{\partial t} = \{ H_{\mbox{\scriptsize tot}}, f\}.
\label{eq:Liouville_eq}
\end{equation}
We define the variables. The position and momentum coordinates for 
the whole system is $\Gamma=(\Gamma_{s},\Gamma_{b})$, 
where $\Gamma_{s}$ and $\Gamma_{b}$ is the coordinates for the systems 
and the baths, respectively. We set $f = \rho\chi\eta$.
$\rho$ and $\chi\eta$ are 
\begin{equation}
\rho = \int d\Gamma_{b} \; f, \;\; \chi\eta = \int d\Gamma_{s} \; f.
\end{equation}
The normalization conditions are
\begin{equation}
\int d\Gamma_{s} \; \rho = 1,\;\; \int d\Gamma_{b}\; \chi\eta = 1. 
\end{equation}
Integrating eq.~(\ref{eq:Liouville_eq}) with respect to the bath coordinates, 
we obtain
\begin{equation}
\frac{\partial \rho}{\partial t} + \{ \rho, H_{s}\} + 
\int d\Gamma_{b} \; \{ f, H_{b}\} + \int d\Gamma_{b}\;\{ f, U\} = 0. 
\end{equation}
The third term is the divergence. Thus, this is zero
if the contribution vanishes sufficiently rapidly toward the surface of 
the system. This point is very similar to the case in Zubarev's derivation. 
Here we omit the term which appears in the use of the Gauss theorem.  
Then, we get 
\begin{equation}
\frac{\partial \rho}{\partial t} + \{ \rho, H_{s}\} + 
\frac{\partial (\rho F_{\alpha})}{\partial p_{\alpha}} = 0.
\label{eq:system_Liouville_eq}
\end{equation}
Here the force $F_{\alpha}$ is 
\begin{equation}
F_{\alpha} = - \int d\Gamma_{b} \chi\eta \frac{\partial U}{\partial q_{\alpha}}.
\end{equation}
This force is not conservative. 
If one set 
\begin{equation}
\frac{\partial F_{\alpha}}{\partial p_{\alpha}} = 
- \int d{\bf S} \cdot{\bf J}_{S}({\bf x};t),
\end{equation}
then one obtains the McLennan-Zubarev steady state distribution $\rho_{ss}$. 
Here ${\bf J}_{S}({\bf x};t)$ is 
\begin{equation}
{\bf J}_{S}({\bf x};t) = 
\beta({\bf x};t) ( 
{\bf J}_{H}(\Gamma,{\bf x}) 
- \mu^{*}{\bf j}(\Gamma,{\bf x}) 
- {\bf u}({\bf x}) \cdot \mathsf{T}(\Gamma,{\bf x})
).
\end{equation}
This is the way of McLennan's derivation. 

\section*{Appendix B \label{AppendixB}} 
\setcounter{section}{2}
\setcounter{equation}{0}

In this appendix, we note a character of the Liouvillian. 
Remember characters of the Liouville equation.
\begin{equation}
\frac{\partial \rho}{\partial t} = {\cal L}\rho,
\end{equation}
where 
\begin{equation}
{\cal L} = \{H, \; \} = 
\sum_{i=1}^{N} \left \{ 
\frac{\partial H}{\partial {\bf q}_{i}}
\cdot
\frac{\partial \; }{\partial {\bf p}_{i}}
-
\frac{\partial H}{\partial {\bf p}_{i}}
\cdot
\frac{\partial \; }{\partial {\bf q}_{i}}
\right \}.
\end{equation}
is the Liouvillian. The formal solution of the distribution is 
\begin{equation}
\rho(\Gamma,t) = e^{{\cal L}t} \rho(\Gamma,0).
\end{equation}
The coordinates is evolved as
\begin{equation}
\Gamma(t) = e^{-{\cal L}t}\Gamma(0).
\end{equation}
The Liouvillian has an anti-linearity.
\begin{equation}
\int d\Gamma \; F(\Gamma)[{\cal L}G(\Gamma)] 
= - \int d\Gamma \; [{\cal L}F(\Gamma)] G(\Gamma).
\end{equation}
This anti-linearity implies the following.
\begin{eqnarray}
\int d\Gamma \; F(\Gamma) G({\cal T}(\Gamma)) 
& = &
\int d\Gamma \; F(\Gamma) {\cal T} G(\Gamma) \nonumber \\
& = &
\int d\Gamma \; G(\Gamma) {\cal T}^{-1} F(\Gamma) \nonumber \\
& = &
\int d\Gamma \; G(\Gamma) F({\cal T}^{-1}(\Gamma)).
\end{eqnarray}
Here we set ${\cal T}=e^{-{\cal L}\tau}$ which is 
a time-evolution operator for the coordinates. 
This identity is used in the text.

\end{document}